\documentclass[pdflatex,sn-mathphys-num]{sn-jnl}

\usepackage{siunitx}
\usepackage{graphicx}%
\usepackage{multirow}%
\usepackage{amsmath,amssymb,amsfonts}%
\usepackage{amsthm}%
\usepackage{mathrsfs}%
\usepackage[title]{appendix}%
\usepackage{xcolor}%
\usepackage{textcomp}%
\usepackage{manyfoot}%
\usepackage{booktabs}%
\usepackage{algorithm}%
\usepackage{algorithmicx}%
\usepackage{algpseudocode}%
\usepackage{listings}%

\theoremstyle{thmstyleone}%
\theoremstyle{thmstyletwo}%

\theoremstyle{thmstylethree}%

\begin{document}

\title[Conceptual Design of a Novel Highly Granular Crystal Electromagnetic Calorimeter for Future Higgs Factories]{Conceptual Design of a Novel Highly Granular Crystal Electromagnetic Calorimeter for Future Higgs Factories}


\author[1,2,3]{\fnm{Baohua} \sur{Qi}}\email{qibh@ihep.ac.cn}

\author[1,4]{\fnm{Fangyi} \sur{Guo}}\email{guofangyi@ihep.ac.cn}

\author[5,6,7]{\fnm{Shu} \sur{Li}}\email{shuli@sjtu.edu.cn}

\author*[1,2,3]{\fnm{Yong} \sur{Liu}}\email{liuyong@ihep.ac.cn}

\author[1,3]{\fnm{Manqi} \sur{Ruan}}\email{manqi.ruan@ihep.ac.cn}

\author[1,3]{\fnm{Weizheng} \sur{Song}}\email{songwz@ihep.ac.cn}

\author[1,3]{\fnm{Shengsen} \sur{Sun}}\email{sunss@ihep.ac.cn}

\author[1,2,3]{\fnm{Yifang} \sur{Wang}}\email{yfwang@ihep.ac.cn}

\author[1,8]{\fnm{Yuexin} \sur{Wang}}\email{wangyuexin@ihep.ac.cn}

\author[5,6,7]{\fnm{Haijun} \sur{Yang}}\email{haijun.yang@sjtu.edu.cn}

\author[1,3]{\fnm{Yang} \sur{Zhang}}\email{zhangyang98@ihep.ac.cn}

\author[6,7]{\fnm{Zhiyu} \sur{Zhao}}\email{zhiyuzhao@sjtu.edu.cn}

\affil[1]{\orgdiv{Institute of High Energy Physics}, \orgname{Chinese Academy of Sciences}, \orgaddress{\street{19B Yuquan Road}, \city{Beijing}, \postcode{100049}, \country{China}}}

\affil[2]{\orgname{State Key Laboratory of Particle Detection and Electronics}, \orgaddress{\street{19B Yuquan Road}, \city{Beijing}, \postcode{100049}, \country{China}}}

\affil[3]{\orgname{University of Chinese Academy of Sciences}, \orgaddress{\street{19A Yuquan Road}, \city{Beijing}, \postcode{100049}, \country{China}}}

\affil[4]{\orgname{China Center of Advanced Science and Technology}, \orgaddress{\street{55 Zhongguancun East Road}, \city{Beijing}, \postcode{100190}, \country{China}}}

\affil[5]{\orgdiv{Institute of Nuclear and Particle Physics, School of Physics and Astronomy}, \orgname{Shanghai Jiao Tong University}, \orgaddress{\street{800 Dongchuan Road}, \city{Shanghai}, \postcode{200240}, \country{China}}}

\affil[6]{\orgdiv{Tsung-Dao Lee Institute}, \orgname{Shanghai Jiao Tong University}, \orgaddress{\street{1 Lisuo Road}, \city{Shanghai}, \postcode{201210}, \country{China}}}

\affil[7]{\orgdiv{National Key Laboratory of Dark Matter Physics, Shanghai Key Laboratory for Particle Physics and Cosmology}, \orgname{Key Laboratory for Particle Astrophysics and Cosmology (MoE)}, \orgaddress{\street{800 Dongchuan Road}, \city{Shanghai}, \postcode{200240}, \country{China}}}

\affil[8]{\orgname{China Spallation Neutron Source Science Center}, \orgaddress{\street{1 Zhongziyuan Road}, \city{Dongguan}, \postcode{523803}, \state{Guangdong}, \country{China}}}


\abstract{Next-generation high-energy electron-positron colliders, operating as Higgs factories, require an unprecedented jet energy resolution for precision measurements of Higgs and Z/W bosons. To address this challenge, a conceptual design is presented for a novel high-granularity crystal electromagnetic calorimeter that combines the superior intrinsic energy resolution of a homogeneous calorimeter with the fine segmentation required for particle-flow reconstruction. The crystal electromagnetic calorimeter design is based on orthogonally arranged long scintillating crystal bars read out by silicon photomultipliers (SiPMs) at both ends. Key design specifications were established through comprehensive simulation studies. Critical technical considerations, including crystal choices, photosensors, electronics, mechanical support, and radiation damage, are discussed. A dedicated digitisation framework was developed to realistically model effects from the crystal, SiPMs, and readout electronics. The performance of a single calorimeter module was evaluated using simulated electron showers. Simulation results for a single module demonstrate an excellent electromagnetic energy resolution of $1.12\%/\sqrt{E(\mathrm{GeV})}\oplus0.22\%$ and an energy linearity within $\pm0.5\%$ for electrons from \qtyrange{3}{100}{\GeV}. The performance significantly exceeds the design requirement of $\leq3\%/\sqrt{E(\mathrm{GeV})}\oplus1\%$. The results establish the feasibility of the proposed high-granularity crystal calorimeter concept and point to a promising pathway toward the precision calorimetry required for future high-energy electron-positron collider experiments.}


\maketitle

\section{Introduction}\label{sec:intro}
Following the landmark discovery of the Higgs boson at the Large Hadron Collider (LHC) in 2012~\cite{the_atlas_collaboration_observation_2012,the_cms_collaboration_observation_2012}, precision measurements of its properties have become a crucial objective for next-generation high-energy collider experiments. The concept of a high-energy electron-positron collider serving as a Higgs factory with high-luminosity and a high-purity collision environment has formed a consolidated strategic consensus in the particle physics community~\cite{european_strategy_group_2020_2020,butler_report_2023}. Several electron-positron collider projects have been proposed as potential Higgs factories, such as the Circular Electron Positron Collider (CEPC)~\cite{the_cepc_study_group_cepc_2024}, the Future Circular Collider electron-electron (FCC-ee)~\cite{the_fcc_collaboration_fcc-ee_2019}, the International Linear Collider (ILC)~\cite{behnke_international_2013} and the Compact Linear Collider (CLIC)~\cite{burrows_multi-tev_2012}, aiming to conduct in-depth studies of the Higgs, Z and W bosons.

Driven by the physics demand for precise Higgs measurements, the large detectors for the future Higgs factories require an unprecedented jet energy resolution to reach an invariant mass resolution better than $4\%$ for the hadronic decays of Higgs and Z/W bosons~\cite{the_cepc_study_group_cepc_2018}. To meet the stringent physics requirements, a novel detector system based on the Particle Flow Approach (PFA)~\cite{brient_calorimetry_2001} demonstrates promising jet reconstruction capabilities. The PFA-oriented detector is designed to leverage the advantages of different sub-detectors and employ the most suitable one for precisely measuring each final-state particle. 

As a critical detector component, PFA calorimeters are favourable for efficient particle reconstruction by making optimal use of available shower information and integrating it with the tracking system, which requires fine segmentation of the detector cells. In addition to measuring energy, they provide three-dimensional spatial and temporal information, allowing for clear imaging of particle showers. With dedicated reconstruction algorithms, high-granularity calorimeters enable the separation and measurement of individual particles within the shower, serving as the fundamental detector hardware for the PFA application.

The high-granularity electromagnetic calorimeters (ECALs) are mostly based on sampling structure, such as the silicon-tungsten electromagnetic calorimeter (SiW ECAL)~\cite{the_calice_collaboration_design_2008} and the scintillator-tungsten electromagnetic calorimeter (ScW ECAL)~\cite{the_calice_collaboration_performance_2014}. Since the sampling design would impose a limitation on the energy resolution (at the level of around $10\%/\sqrt{E(\mathrm{GeV})}$), a homogeneous ECAL with an excellent intrinsic energy resolution and a sufficiently low detection limit for electromagnetic (EM) showers will have significant advantages for the physics programmes at future lepton collider experiments.

Building upon the initial concept proposed by the CEPC calorimeter working group~\citep{liu_high-granularity_2020}, a novel high-granularity crystal electromagnetic calorimeter has been under active development, with significant progress in research and development achieved~\citep{qi_rd_2022,qi_development_2025}. The design features finely segmented crystal cells compatible with PFA, along with a homogeneous structure to achieve an optimal EM energy resolution of $\leq3\%/\sqrt{E(\mathrm{GeV})}\oplus1\%$. The high-granularity crystal ECAL enables the precise identification and measurement of EM showers, thereby effectively supporting tasks such as Bremsstrahlung energy loss correction and Higgs boson recoil mass reconstruction, and leading to improved Boson Mass Resolution (BMR) for $H\to\gamma\gamma$ and $Z\to e^{+}e^{-}$ processes. Significant improvements in sensitivity to low-energy particles with crystal enhance precision measurements of $\pi^{0}$, thus providing unique opportunities for flavour physics~\cite{wang_prospects_2022} and furthering the exploration of new physics beyond the Standard Model~\cite{fang_physics_2023}.

This paper provides a comprehensive introduction to the conceptual design of the high-granularity crystal ECAL. Section~\ref{sec:Design} elaborates on the detailed design schemes. Section~\ref{sec:Specification} outlines the requirements for key design specifications. Section~\ref{sec:Consideration} discusses the crucial factors to be considered during the detector realisation phase. Section~\ref{sec:Performance} highlights the studies on the module-level performance, followed by the summary and prospects in Section~\ref{sec:Summary}.

\section{Conceptual Design}\label{sec:Design}
The high-granularity crystal ECAL combines advanced hardware with sophisticated software to meet the requirements of future lepton colliders. Featuring a design based on finely segmented crystals, the system demands an innovative geometric layout as well as a dedicated particle-flow reconstruction software. The following sections elaborate on the conceptual design of this system.

\subsection{Module Design}\label{subsec:ModuleDesign}
The design of the high-granularity crystal ECAL generally follows two conceptual schemes. The first and most straightforward approach involves segmenting the ECAL modules three-dimensionally into a large number of short crystal units, which are inherently suitable for PFA. However, this design presents several challenges: the large number of channels results in higher costs and increased cooling demands. The inter-layer components, such as readout electronics, cooling plates, and mechanical structures, introduce substantial insensitive material, degrading the detector's performance. To significantly reduce the number of readout channels and material budget between sensitive layers, while maintaining a high-granularity layout, another design concept with long crystal bars has been proposed as the baseline option.

\begin{figure}[htbp]
\centering
\includegraphics[width=0.7\textwidth]{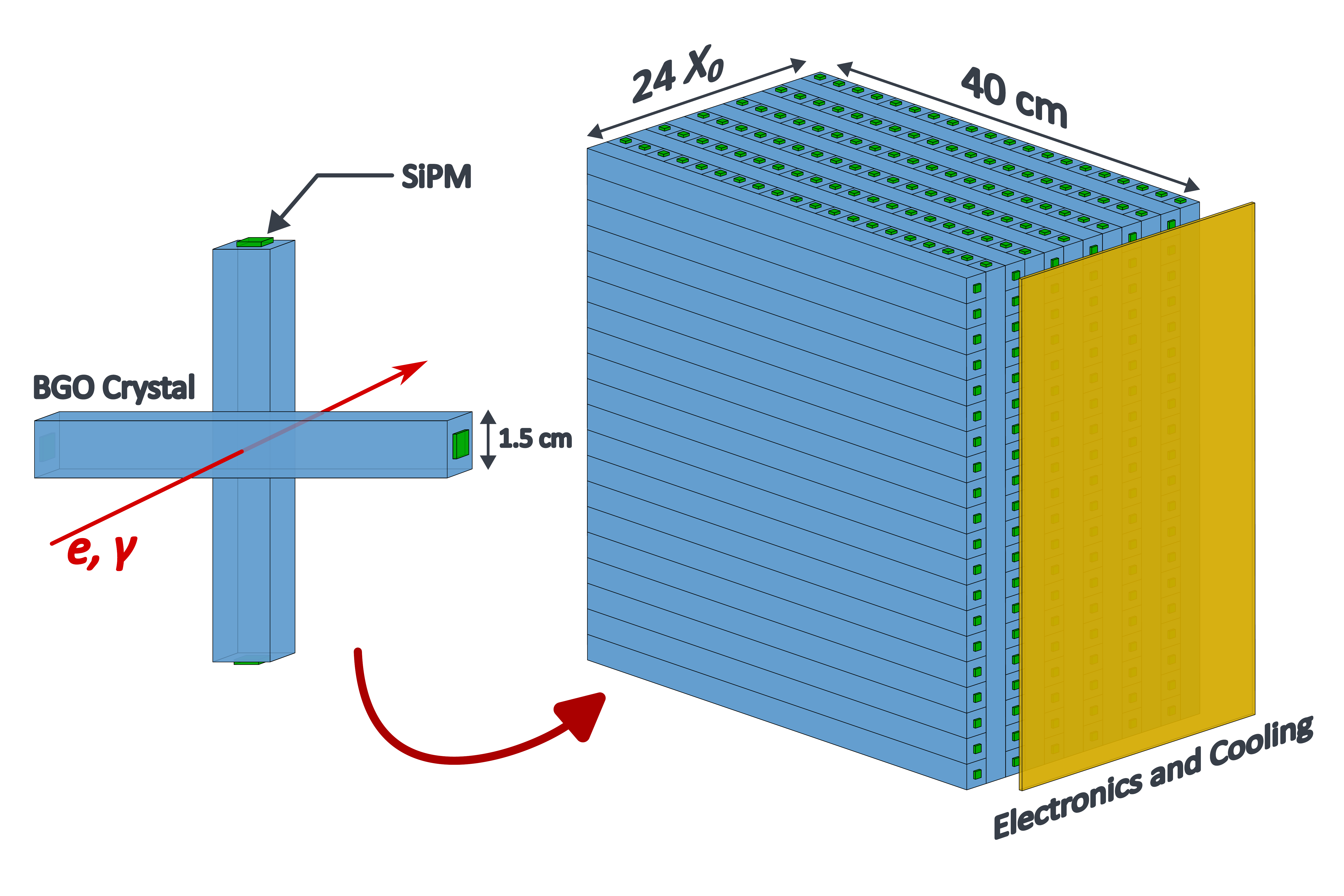}
\caption{The schematics of the typical module of the high-granularity crystal ECAL. Long scintillating crystal bars are arranged orthogonally between adjacent layers to provide longitudinal segmentation and effective transverse granularity. Each crystal bar is wrapped in a reflective foil to enhance light collection efficiency and response uniformity, and is read out by two SiPMs, one coupled to each end. The front-end electronics and the cooling system are placed on the four sides of the module.}\label{fig:CrystalECALModuleDesign}
\end{figure}

As illustrated in Figure~\ref{fig:CrystalECALModuleDesign}, a typical ECAL module consists of long crystal bars arranged orthogonally between adjacent layers. This configuration enables fine longitudinal segmentation, while the transverse granularity is achieved by combining information from pairs of neighbouring layers. Each module features an area of \qtyproduct[product-units=power]{40x40}{\cm} and a thickness of 24 radiation length ($X_{0}$) for effective containment of EM showers. High-density inorganic scintillating crystals, such as Bismuth Germanate (BGO), are employed as the sensitive material, each wrapped in reflective foil. Scintillation light from each crystal bar is read out by two Silicon Photomultipliers (SiPMs), one mounted at each end, to improve response uniformity. The front-end electronics and the cooling system are compactly integrated along all four sides of the modules.

To ensure good separation of adjacent particle showers for PFA, the longitudinal segmentation within the modules should be on the order of the radiation length, while the transverse size should be on the order of the Moli\`ere radius ($R_{\mathrm{M}}$). As an optimised result considering performance, fabrication feasibility, and detector modularity, the standard crystal size is chosen as \qtyproduct[product-units=power]{1.5x1.5x40}{\cm}. Thus, a typical module comprises 18 layers and 486 crystal bars, with 972 SiPM readout channels providing energy, position, and timing information.

\subsection{Full ECAL Design}\label{subsec:FullECALDesign}
The overall design of the full crystal ECAL adopts a cylindrical geometry, including a barrel sector and two endcap sectors, necessitating the careful optimisation of the module shape and layout. This optimisation aims to achieve complete solid-angle coverage for detector hermeticity and to prevent energy leakage in the inter-module regions.

\begin{figure}[htbp]
\centering
\includegraphics[width=0.48\textwidth]{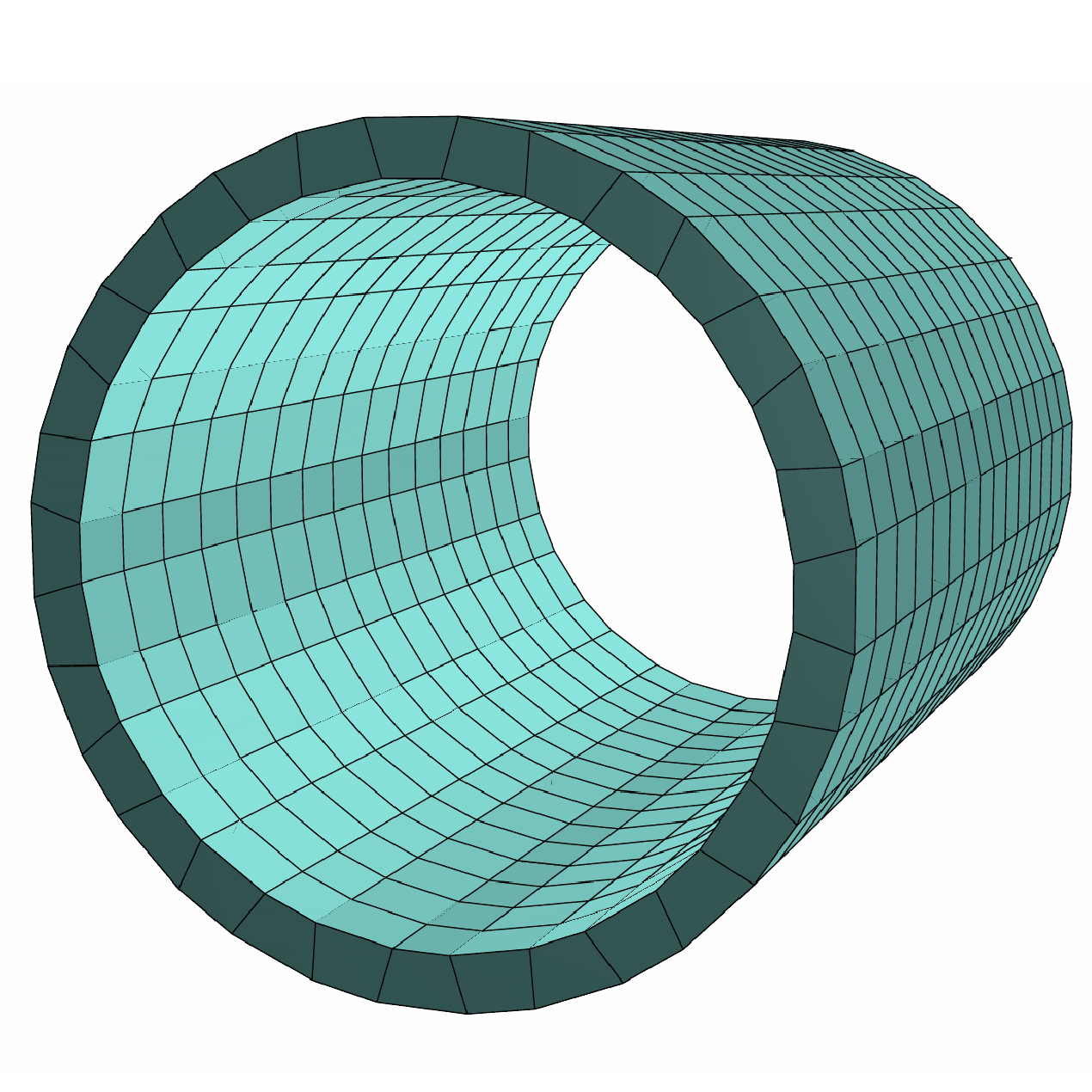}
\includegraphics[width=0.48\textwidth]{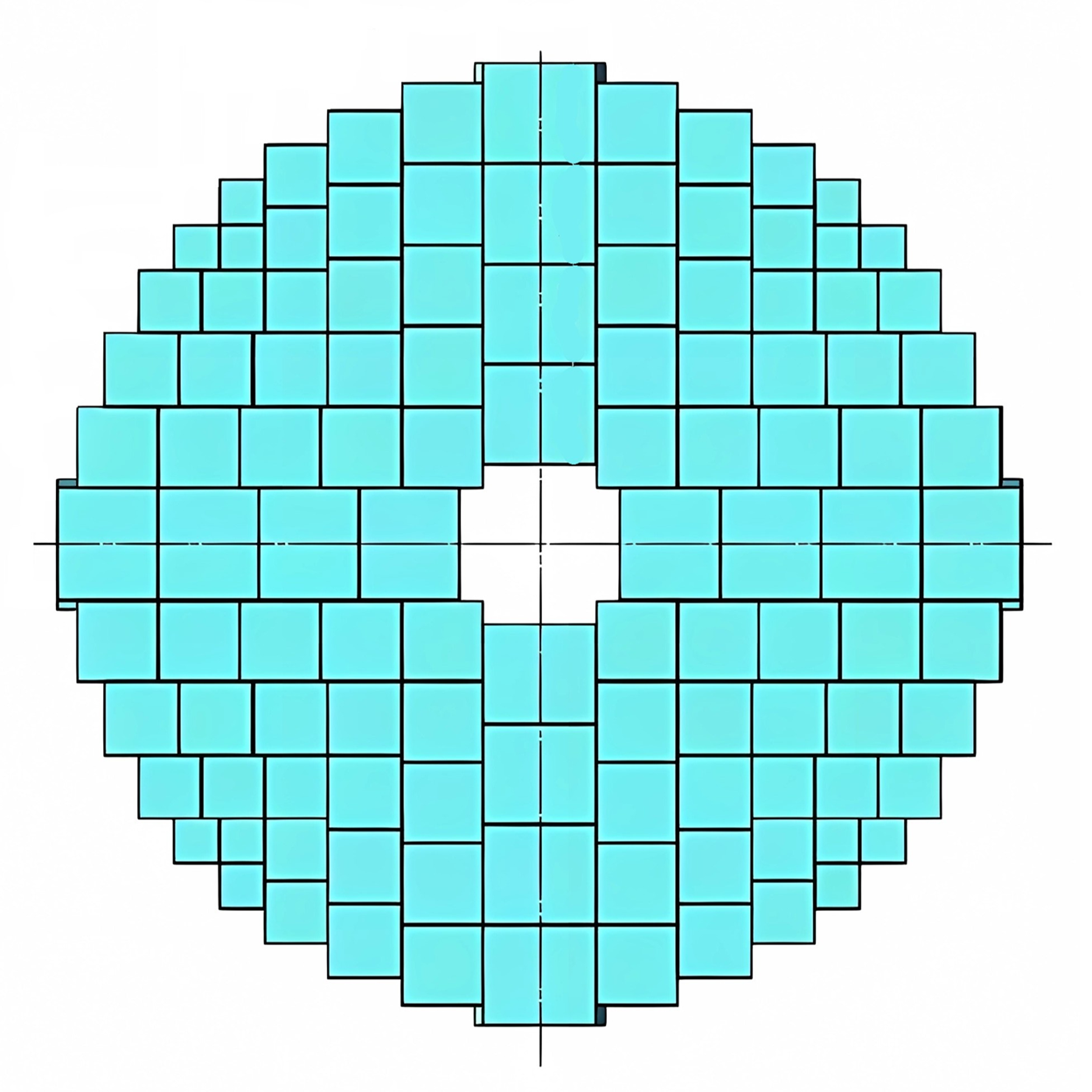}
\caption{The schematics of the preliminary modular geometry design of the high-granularity crystal ECAL for the CEPC. The cylindrical ECAL barrel (left) comprises 480 modules, and the disk-shaped endcap (right) contains 112 modules.}\label{fig:FullCrystalECALDesign}
\end{figure}

Figure~\ref{fig:FullCrystalECALDesign} illustrates the overall geometry of the ECAL barrel and endcaps, using the CEPC reference detector as a representative case. The ECAL barrel has an axial length of \qty{5800}{\mm}, with inner and outer radii of \qty{1830}{\mm} and \qty{2130}{\mm}, respectively. To achieve a well-modularised structure, the barrel is divided into 15 rings, each consisting of 32 sectors. Given that the front-end electronics, cooling system and mechanical support are integrated around the modules, the overall geometric layout poses a key challenge. To avoid any geometric gaps pointing toward the interaction point, the modules are designed in two complementary trapezoidal shapes, and the tilt angle between them has been carefully optimised~\cite{song_novel_2026}. The endcaps, with an inner radius of \qty{350}{\mm} and an outer radius of \qty{2130}{\mm}, feature a more complex design and require several different module types to maximise the detector coverage. The preliminary detector design parameters of the high-granularity crystal ECAL for the CEPC are listed in Table~\ref{tab:CrystalECALDetectorParameters}. A more detailed description of the geometry design will be presented in a dedicated publication.

\begin{table}[htbp]
\caption{Preliminary detector parameters of the high-granularity crystal ECAL for the CEPC.}  
\label{tab:CrystalECALDetectorParameters}
\begin{tabular}{ll}
\toprule
\textbf{Parameters} & \textbf{Values} \\
\midrule
Barrel Length & \qty{5800}{\mm} \\
Barrel Inner Radius & \qty{1830}{\mm} \\
Barrel Outer Radius & \qty{2130}{\mm} \\
Barrel Modularity & 32 modules $\times$ 15 rings = 480 modules \\
Endcap Inner Radius & \qty{350}{\mm} \\
Endcap Outer Radius & \qty{2130}{\mm} \\
Endcap Modularity & 112 modules $\times$ 2 sectors = 224 modules \\
ECAL Thickness & $\sim24\,X_{0}$ \\
Total Crystal Material Budget & $\sim\qty{24}{\m^3}$ ($\sim171$ tons of BGO) \\
Total Readout Channels & $\sim0.57$M \\
\botrule
\end{tabular}
\end{table}

\subsection{Particle-flow Software}\label{subsec:PFASoftware}
The PFA reconstruction software is a critical subject in the development of the high-granularity calorimeters. Compared to sampling ECALs, the homogeneous crystal ECAL features a larger Moli\`ere radius and a lower ratio of nuclear interaction length ($\lambda_{\mathrm{I}}$) to the radiation length, resulting in increased overlap of showers from different particles within jets. Furthermore, the unique geometry design with orthogonally arranged long crystal bars introduces ambiguities. When two or more particles hit the calorimeter simultaneously, multiple combinations of hit positions can arise, leading to the so-called ``ghost hit'' problem. Addressing these challenges in pattern recognition requires the development of dedicated reconstruction algorithms.

To fully exploit the physics potential of the high-granularity crystal ECAL, a dedicated reconstruction software, CyberPFA (CrYstal Bar Ecal Reconstruction PFA), has been developed for the CEPC and implemented within the CEPCSW~\cite{the_cepc_study_group_cepcsw_2020} software framework. The algorithm employs a multi-stage clustering strategy to handle the complicated hit environments. 

Initially, the algorithm clusters neighbouring hits globally to identify local energy deposition maxima, thus realising preliminary separation of particle showers. Subsequently, for different particle types, it integrates algorithms such as the Hough transformation ($\gamma$), track matching (charged) and cone clustering (neutral) to accurately identify the shower core and reconstruct the topological structure of particle showers~\cite{zhang_photon_2026}. To precisely reconstruct overlapping EM showers, the algorithm performs energy splitting based on the lateral shower profile around the shower core. As for the ghost hit issue, the hit ambiguity can be effectively resolved by using track extrapolation, information from neighbouring modules, and longitudinal shower profiles. Moreover, timing information also holds potential for further application in the reconstruction process, enabling five-dimensional calorimetry. A dedicated paper detailing the PFA algorithm developed specifically for the crystal ECAL will be published separately.

\section{Key Specifications}\label{sec:Specification}
The design specifications of the high-granularity crystal ECAL are defined based on the following performance requirements: (a) an optimal EM energy resolution of $\leq3\%/\sqrt{E(\mathrm{GeV})}\oplus1\%$, and (b) compatibility with PFA, which requires efficient separation of individual particle showers to achieve a benchmark performance of BMR better than 4\% for jet final states. Based on the preliminary research, a series of key design specifications is provided in Table~\ref{tab:CrystalECALSpecifications} and will be discussed in the following content. Further validation and refinement efforts are ongoing.

\begin{table}[htbp]
\caption{Key design specifications of the high-granularity crystal ECAL.}  
\label{tab:CrystalECALSpecifications} 
\begin{tabular}{lll}
\toprule
\textbf{Parameters} & \textbf{Values} & \textbf{Remarks} \\
\midrule
MIP Response & \qty{300}{\text{p.e./MIP}} & For optimal EM energy resolution \\
Energy Threshold & \qty{0.1}{\text{MIP}} & To balance resolution and SiPM noise \\
Crystal Response Uniformity & $<1\%$ &  For consistent response of the crystal \\
Dynamic Range & 0.1 to \qty{3000}{\text{MIP}} & The response range per crystal unit \\
Timing Resolution & \qty{0.5}{\ns} & For MIP signals \\
Temperature Gradient & $\leq\qty{6}{\K}$ & For uniform response and noise control \\
\botrule
\end{tabular}
\end{table}

\subsection{MIP Response and Energy Threshold}\label{subsec:MIPandThreshold}
The light output of the crystal, represented in terms of the photon-equivalent (p.e.) per Minimum Ionising Particle (MIP), dominates the stochastic term of the energy resolution. Meanwhile, the energy threshold directly determines the sensitivity to low-energy signals, thereby impacting the energy resolution.

To quantify the design requirements for MIP response and energy threshold, a Geant4~\cite{agostinelli_geant4simulation_2003} simulation study has been conducted with a single ECAL module. The study systematically investigated the EM energy resolution across different parameter configurations, using electrons with energies from \qty{0.1}{\GeV} to \qty{80}{\GeV}.

\begin{figure}[htbp]
\centering
\includegraphics[width=0.7\textwidth]{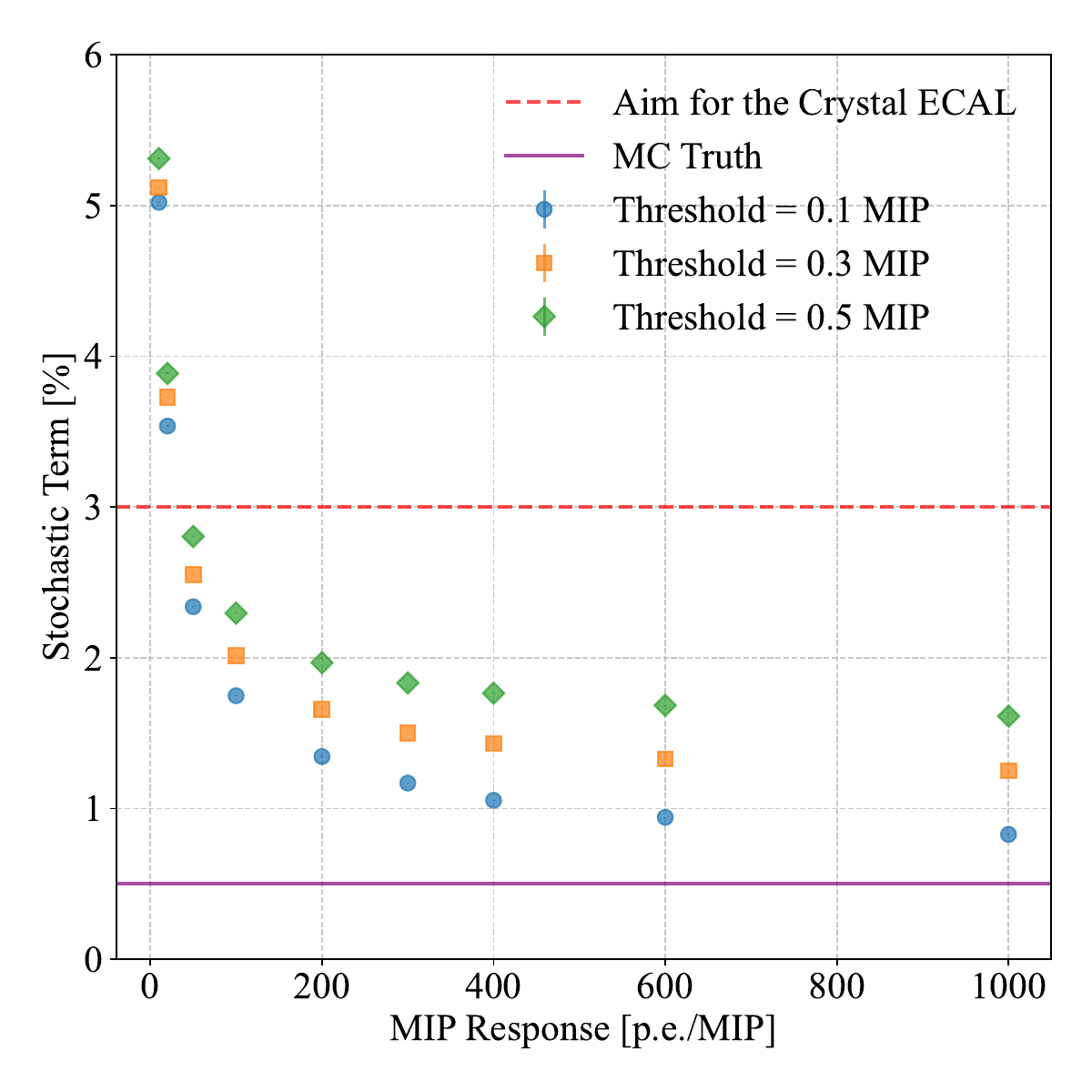}
\caption{Dependence of the stochastic term in EM energy resolution on MIP response and energy threshold per crystal. The purple line indicates the best resolution obtained from the original energy deposition. The red dashed line represents the target upper limit for the stochastic term of the crystal ECAL resolution.}\label{fig:ResolutionVSMIPresponseEnergyThreshold}
\end{figure}

Figure~\ref{fig:ResolutionVSMIPresponseEnergyThreshold} illustrates the dependence of the stochastic term of the EM energy resolution on both the MIP response and the energy threshold. As shown, photon statistics constitute the dominant limiting factor for energy resolution at low crystal MIP responses, while the performance improvement at high response values gradually saturates. Considering the additional contributions from upstream materials and electronics, a sufficient MIP response must be reserved. However, an excessively high MIP response would have a significant impact on the dynamic range of the SiPMs and readout electronics. Balancing these considerations, the design target for the MIP response of the crystal unit is optimised to \qty{300}{\text{p.e./MIP}}, corresponding to an energy deposition of approximately \qty{13.35}{\MeV} in a \qty{1.5}{\cm}-thick BGO crystal.

The energy threshold is particularly critical for low-energy measurements. To enable efficient reconstruction of low-energy particles while accounting for SiPM noise characteristics, the energy threshold for a single crystal bar can be set to \qty{0.1}{\text{MIP}}, corresponding to a per-SiPM threshold of \qty{0.05}{\text{MIP}}, which is achievable using low noise rate and low inter-pixel crosstalk SiPM products.

\subsection{Crystal Response Uniformity}\label{subsec:CrystalUniformity}
In the crystal ECAL that uses long crystal bars as the basic detection units, variations in particle hit positions along the bar may lead to differences in crystal response. This effect results from photon transportation processes such as reflection and absorption. When the intrinsic light attenuation of the crystal is significant, particles incident near the centre exhibit a lower response compared to those hitting near the ends. If this non-uniformity is not effectively controlled, it can result in position-dependent energy reconstruction, seriously degrading the detector's performance.

\begin{figure}[htbp]
\centering
\includegraphics[width=0.7\textwidth]{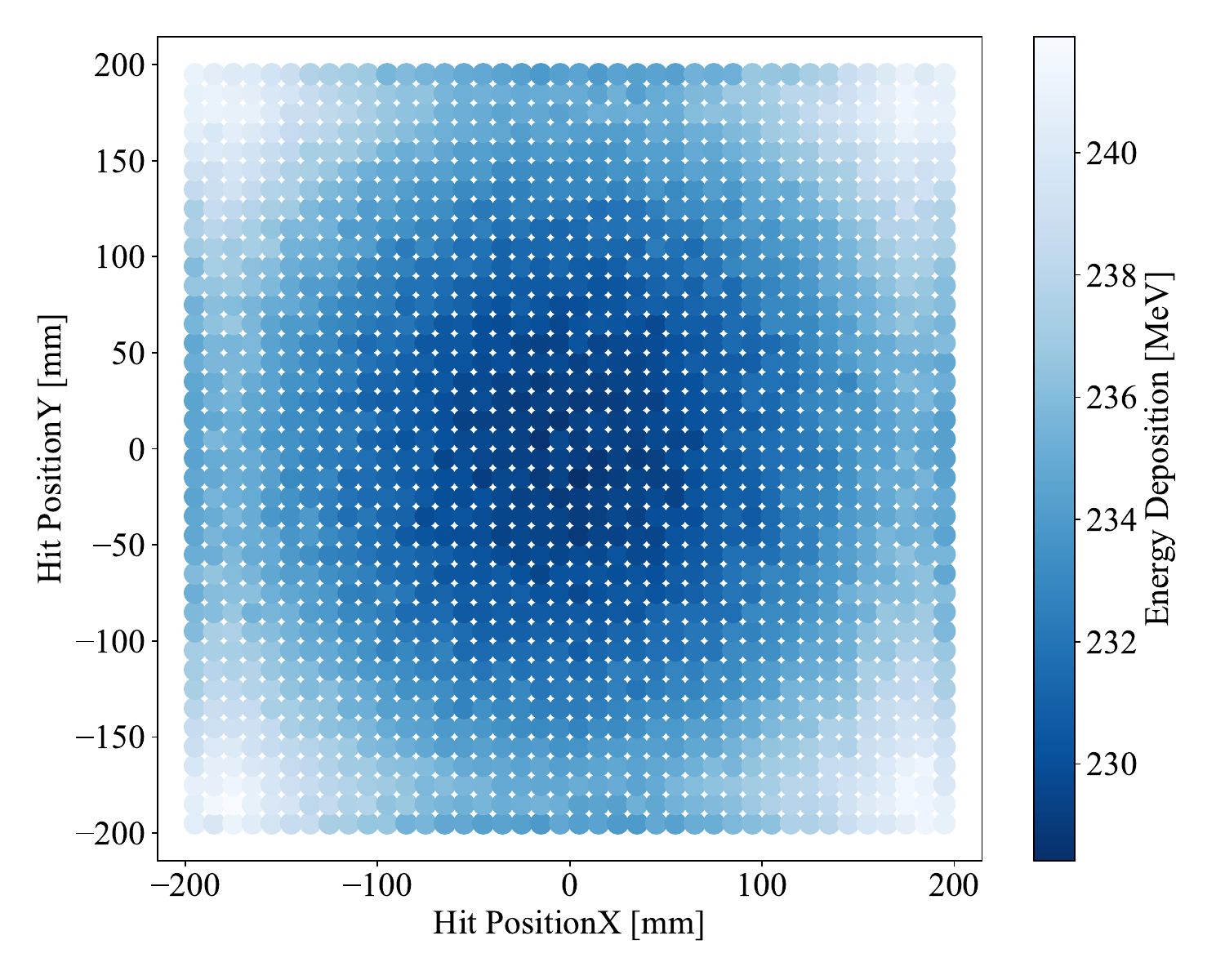}
\caption{A two-dimensional response map of a crystal ECAL module for perpendicularly incident muons at different positions on the module front face. The response is highest at the four corners, followed by the edges, with the lowest response observed in the central region.}\label{fig:Module2DUniformity}
\end{figure}

To study the impact of response non-uniformity, a simplified model of the long crystal bar response can be established by accounting for the exponential attenuation of light in the crystal. Based on this, Geant4 simulations were performed on a single module to examine the response under specified non-uniform conditions. Figure~\ref{fig:Module2DUniformity} shows the reconstructed MIP energy for muons incident perpendicularly at different positions on the module surface, assuming a 10\% lower response at the crystal centre than at the ends. It can be observed that the response is the highest in the four corner regions closest to the SiPM mounting positions, while the central region exhibits the lowest response. This two-dimensional distribution will introduce bias in energy reconstruction.

\begin{figure}[htbp]
\centering
\includegraphics[width=0.7\textwidth]{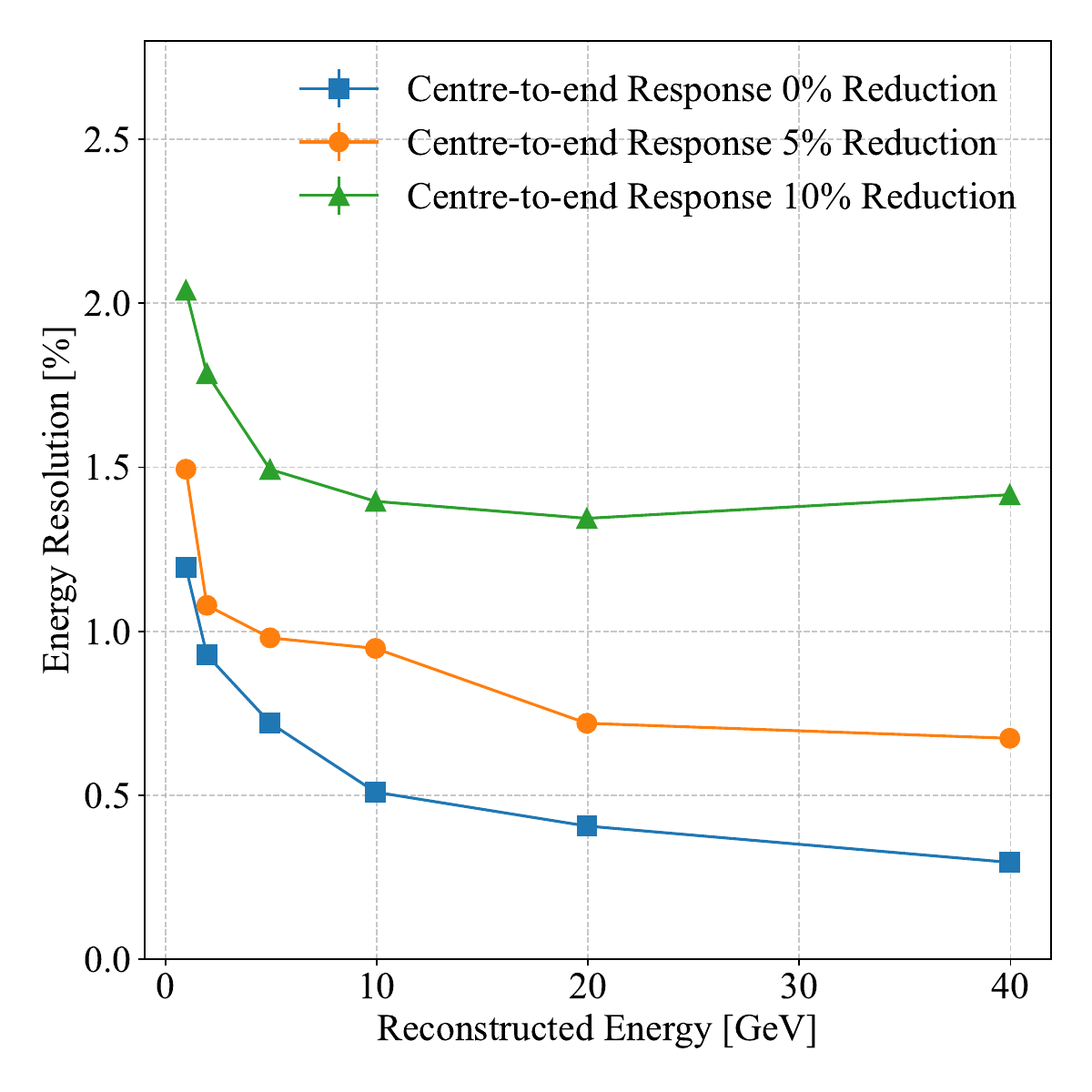}
\caption{Energy resolution under different degrees of crystal centre-to-end response reduction. The study employs nine crystal ECAL modules, with electrons randomly incident on the central one.}\label{fig:ResolutionVSUniformity}
\end{figure}

Figure~\ref{fig:ResolutionVSUniformity} shows the simulated EM energy resolution for electrons randomly incident on the surface of a single module, under varying levels of centre-to-end response reduction. The results indicate that the energy resolution degrades significantly as the crystal response non-uniformity increases. Therefore, to ensure reliable energy resolution and accurate energy reconstruction, the centre-to-edge response variation should be maintained at a level well below 5\%.

In practice, factors such as intrinsic light yield variations, crystal polishing, wrapping, and SiPM coupling can all affect response uniformity. Measurement results of \qty{40}{\cm} long BGO crystal bars with radioactive sources or particle beams typically show fluctuations in the position-dependent response. Generally, the measured response uniformity can be quantified as the ratio of the standard deviation to the mean value. Under this definition, the requirement that the centre-to-edge response drop be well below 5\% corresponds approximately to a measured uniformity better than 1\%. Furthermore, maintaining a consistent response across a large number of crystal bars is also essential. It is therefore necessary to establish a high-precision crystal calibration procedure to characterise the crystal response.

\subsection{Dynamic Range}\label{subsec:DynamicRange}

\begin{figure}[htbp]
\centering
\includegraphics[width=0.48\textwidth]{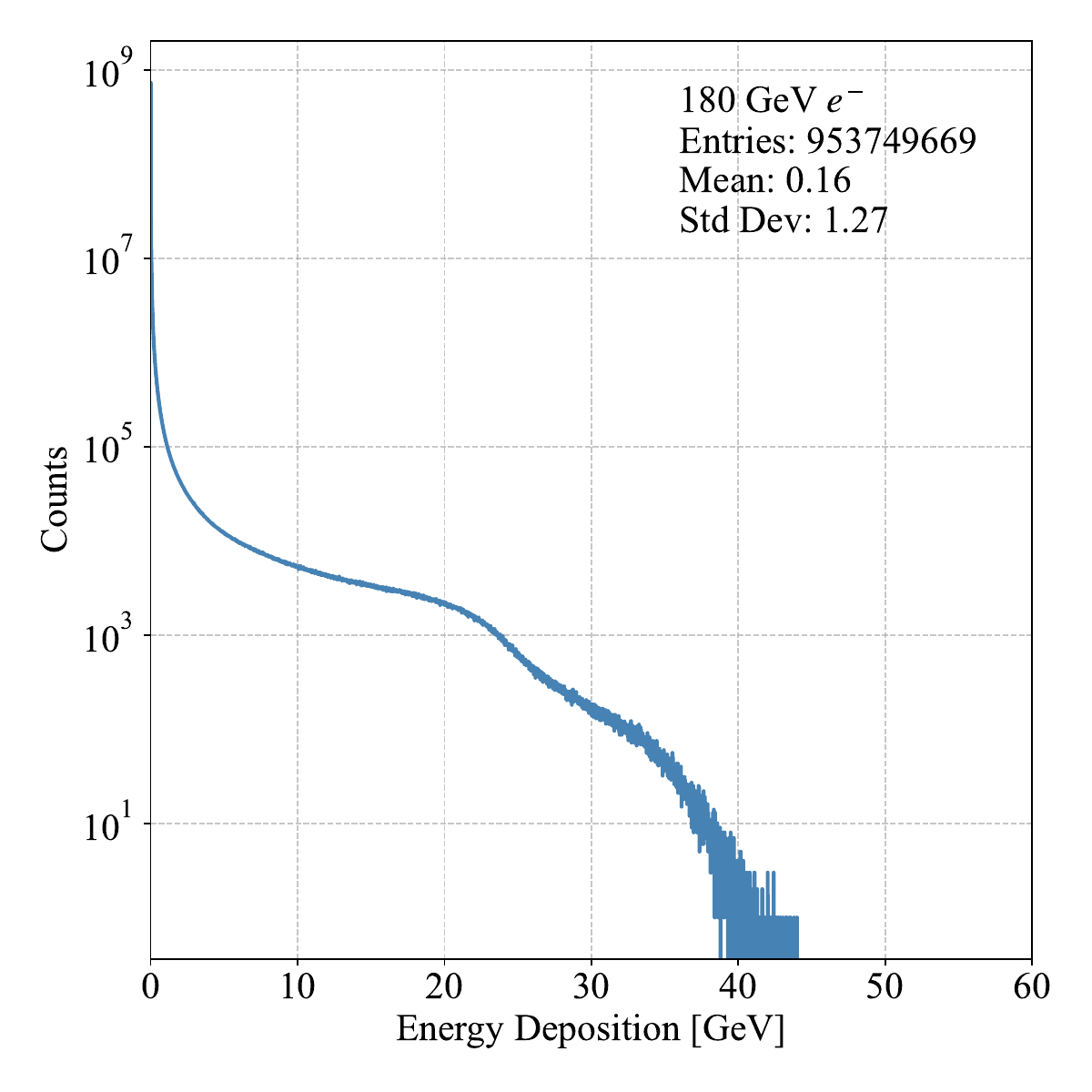}
\includegraphics[width=0.48\textwidth]{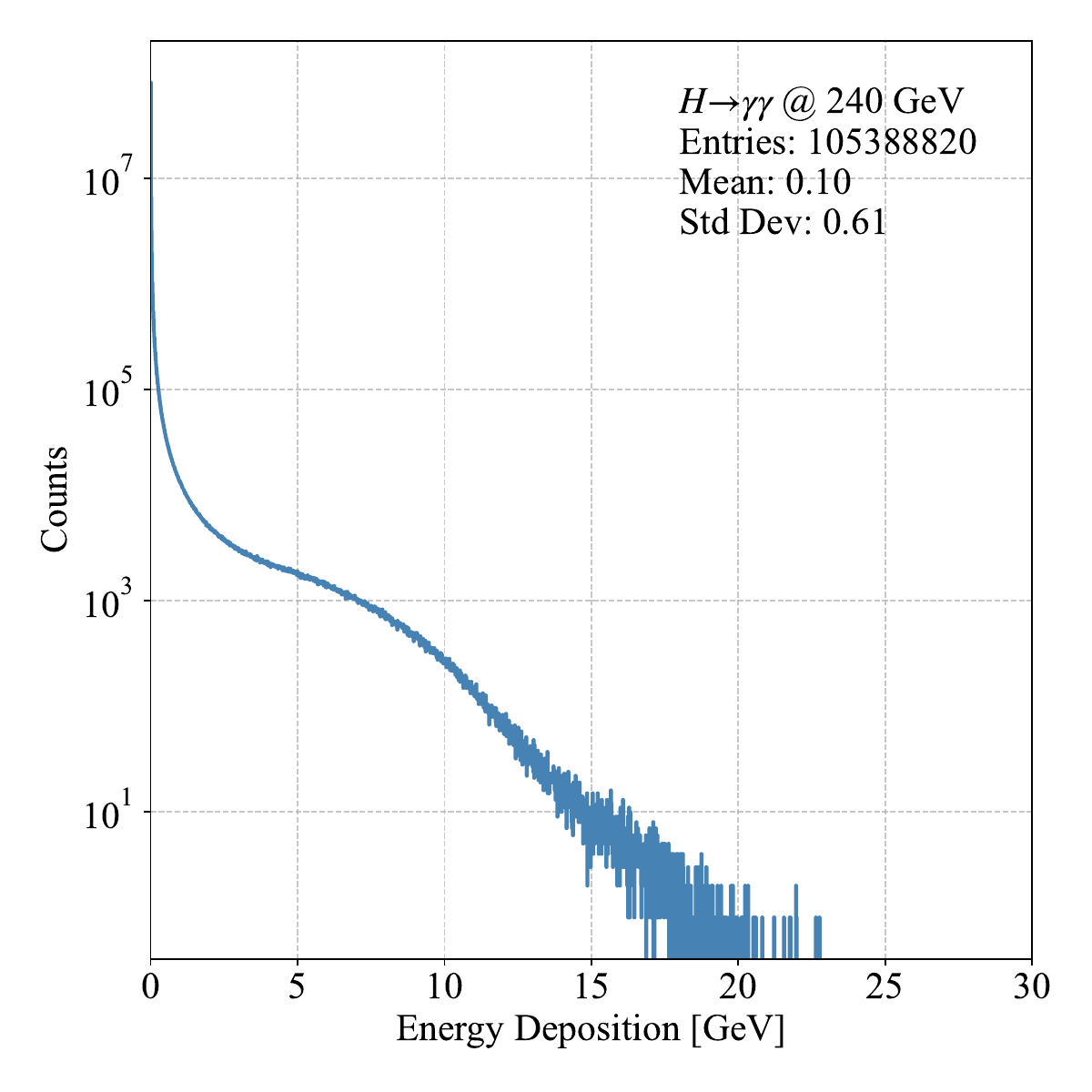}
\caption{Distribution of energy deposition in each crystal cell from \qty{180}{GeV} electron showers (left) and high-energy photon showers in the Higgs boson di-photon decay events (right). The majority of the responses are below \qty{40}{GeV} and \qty{20}{GeV}, respectively.}\label{fig:CrystalHitEnergyElec180GeVAndHtoGG}
\end{figure}

Dynamic range poses a key technical challenge in the design of a crystal ECAL. In future colliders such as the CEPC, the $H\to\gamma\gamma$ decay channel serves as an important probe for Higgs boson measurements, producing high-energy photons that demand accurate energy reconstruction. Meanwhile, in the top-quark operating mode at a centre-of-mass energy of \qty{360}{GeV}, Bhabha scattering from electron–positron collisions can generate electromagnetic showers with energies up to \qty{180}{GeV}. While such high-energy electrons correspond to a relatively small subset of physics processes, they establish an important safety margin for the detector’s upper energy boundary. Without sufficient dynamic range, these high-energy events would lead to significant signal saturation in the crystal ECAL, degrading overall detector performance. On the other end, the detector must also maintain high sensitivity to signals as low as \qty{0.1}{\text{MIP}} to capture weak signatures from specific low-energy processes. Thus, the crystal ECAL is required to operate reliably across an energy range covering five orders of magnitude.

The dynamic range requirement has been evaluated through simulations of EM showers from \qty{180}{GeV} electrons and from the di-photon decays of Higgs bosons. Figure~\ref{fig:CrystalHitEnergyElec180GeVAndHtoGG} shows the distributions of energy deposited in individual crystal cells. In both cases, the energy deposition is highly concentrated in the low-energy region, with most values well below the recorded maxima. For \qty{180}{GeV} electrons, the majority of signals are below \qty{40}{GeV} (approximately \qty{3000}{\text{MIP}}). For the Higgs di-photon channel, most crystal responses are under \qty{20}{GeV} (about \qty{1500}{\text{MIP}}).

Based on these studies, the upper limit of the required dynamic range is set to \qty{3000}{\text{MIP}}, primarily to accommodate the high-energy EM showers expected in the top-quark mode, while also providing a safety margin. The lower limit is defined as \qty{0.1}{\text{MIP}} to ensure good sensitivity to low-energy signals.

\subsection{Timing Resolution}\label{subsec:TimingResolution}

Timing information provided by the crystal ECAL offers valuable inputs to multiple stages of the particle-flow reconstruction, such as clustering, particle identification, and track-cluster matching, thereby potentially improving the overall reconstruction performance. To determine the optimal time resolution achievable by individual crystal bars, the intrinsic timing characteristics have been analysed with Geant4 optical simulations. In these simulations, muons with an energy of \qty{1}{\GeV} incident on the crystal perpendicularly, generating both Cherenkov and scintillation photons. These photons experience multiple reflections within the crystal and at the reflector before being detected by SiPMs.

\begin{figure}[htbp]
\centering
\includegraphics[width=0.7\textwidth]{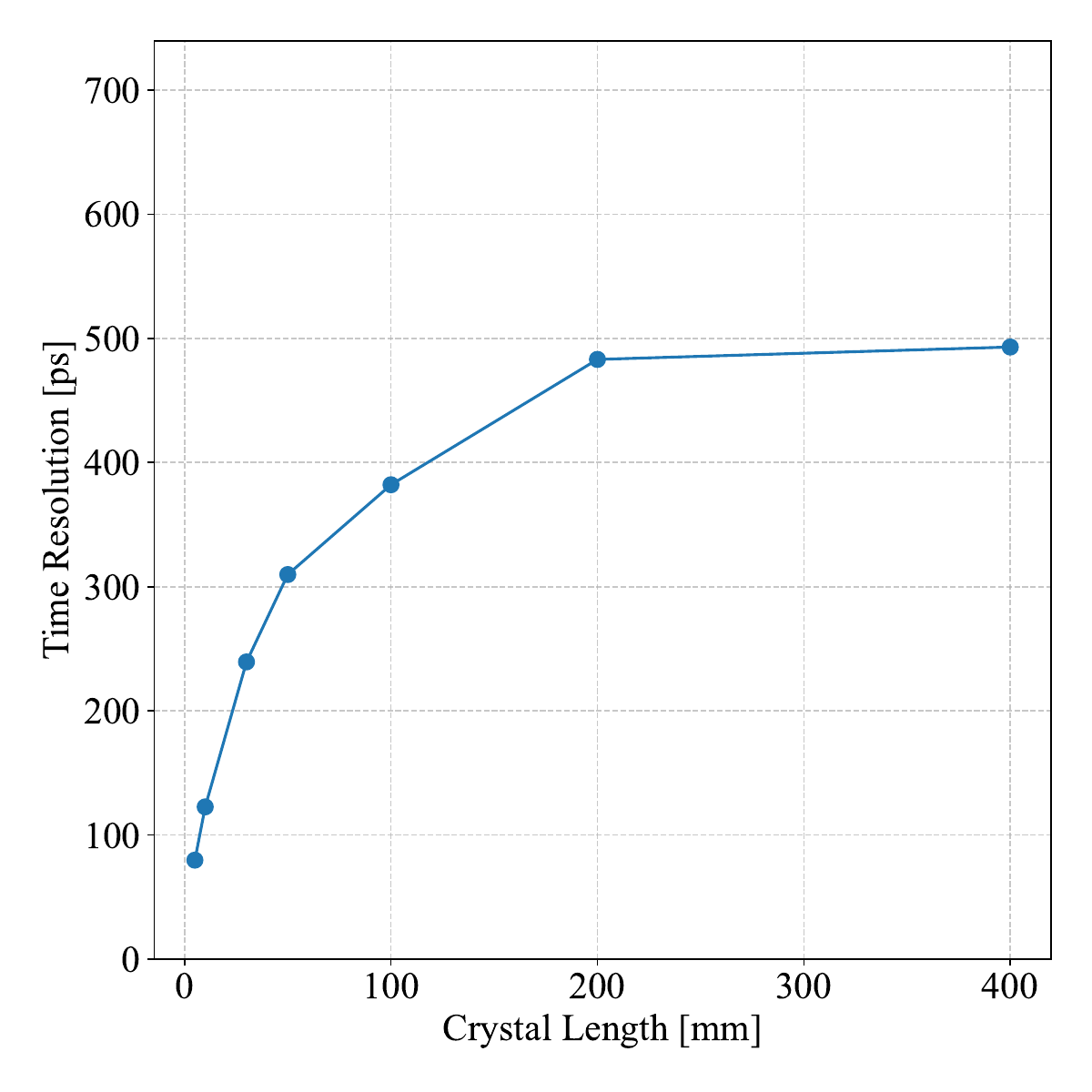}
\caption{Time resolution of BGO crystals of different lengths from optical simulations. The resolution is defined as the spread in the distribution of the sum of the absolute arrival times of the first photons detected at both ends of a single crystal bar.}\label{fig:CrystalBarTimeResolutionVSLength}
\end{figure}

The time resolution of a single crystal bar is defined as the spread of the distribution formed by the sum of the arrival times of the first detected photons at two ends of a crystal. This resolution varies with crystal length due to differences in the path and the number of reflections that photons undergo during propagation. Figure~\ref{fig:CrystalBarTimeResolutionVSLength} presents the simulated time resolution for BGO crystals of various lengths. For a \qty{40}{\cm} long crystal bar, a resolution of approximately \qty{0.5}{\ns} is achieved, representing the best performance attainable by a typical crystal-SiPM unit under single-MIP energy deposition.

Since the timing performance is largely dominated by the crystal itself, with relatively small contributions from other detector effects (e.g., SiPM and readout electronics), the time resolution requirement for a single crystal bar is preliminarily set at \qty{0.5}{\ns} for MIP signals.

\subsection{Temperature Gradient}\label{subsec:TemperatureGradient}
The temperature gradient is a crucial specification for the crystal ECAL, given the pronounced temperature dependence of both the scintillating crystals and the SiPMs. For example, the intrinsic light yield of BGO crystals exhibits a negative temperature dependence, while the gain of the SiPMs also decreases with temperature. Conversely, the SiPM noise rises exponentially with temperature. These effects directly degrade the signal-to-noise ratio and influence the energy measurement of the detector.

\begin{figure}[htbp]
\centering
\includegraphics[width=0.7\textwidth]{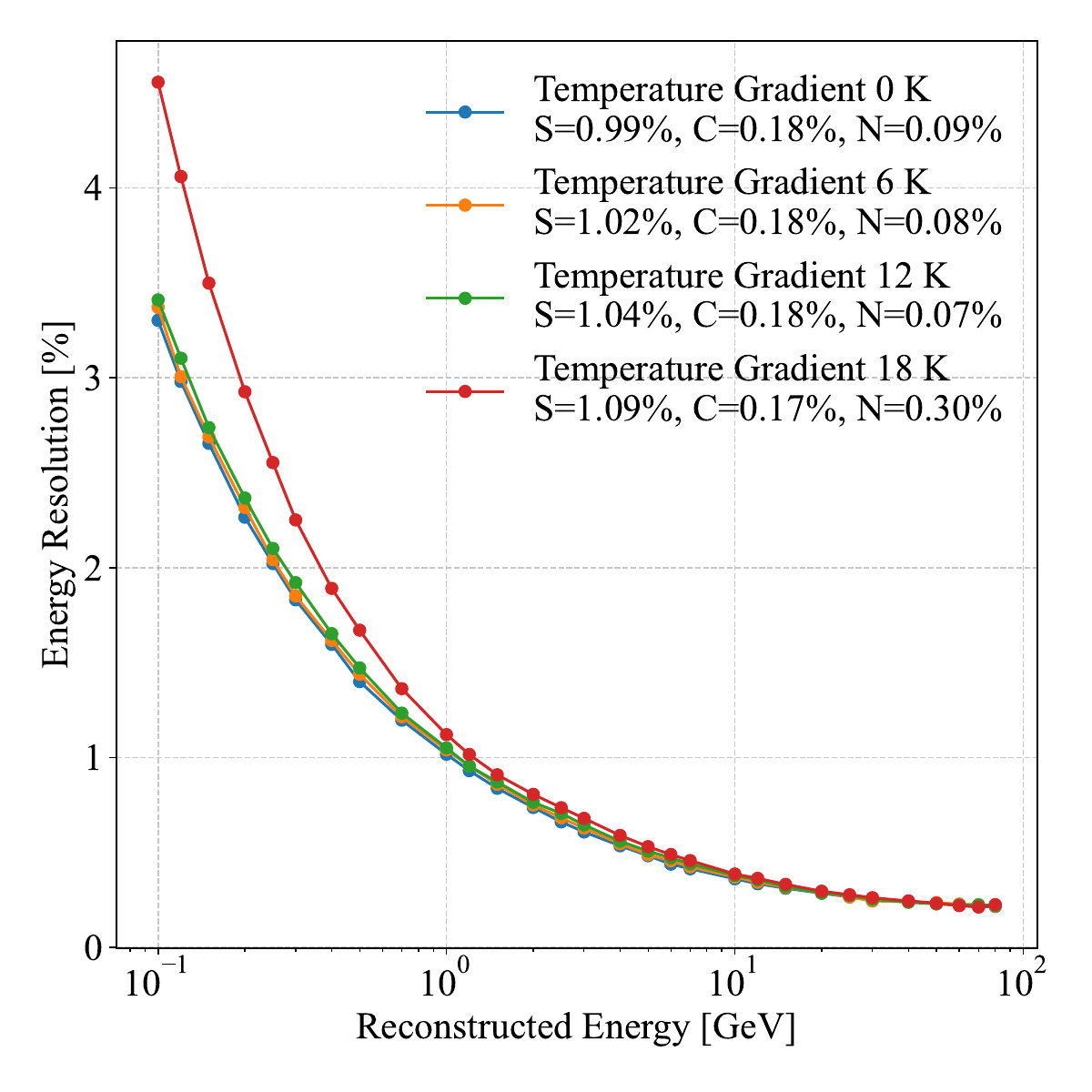}
\caption{Energy resolution versus the reconstructed energy for different temperature gradients of the crystal ECAL. The fitted stochastic term $S$, constant term $C$, and noise term $N$ are presented for each gradient condition.}\label{fig:PerformanceVSTemperatureGradient}
\end{figure}

To study the temperature's impact on performance, the thermal distribution within the full detector is modelled using a linear gradient that decreases from the inner to outer layers of each module. The temperature coefficients of the crystals and SiPMs were implemented into the simulations. As shown in Figure~\ref{fig:PerformanceVSTemperatureGradient}, simulation results indicate that while the constant term of the energy resolution remains largely unaffected, the stochastic term rises with temperature variation. The noise term remains stable under small gradient values but increases notably with larger gradients, especially as more readout channels contribute noise. Analysis shows that temperature gradients up to approximately \qty{6}{\K} have a negligible impact on overall ECAL energy resolution. Therefore, maintaining the temperature gradient within the ECAL structure at $\leq\qty{6}{\K}$ is set as a design requirement, balancing detector performance and engineering feasibility.

Beyond the temperature gradient, maintaining local temperature stability is equally critical to prevent performance fluctuations and calibration drifts over time. For reference, crystal calorimeters such as the CMS ECAL operate with the local temperature stability on the order of \qty{0.05}{\degreeCelsius}~\cite{the_cms_collaboration_cms_2008}. Achieving a uniform and stable thermal environment requires an advanced cooling design with real-time monitoring and active feedback control, taking into account the power consumption, occupancy, and cooling design of the crystal ECAL, balancing detector performance and engineering feasibility.

\section{Design Considerations}\label{sec:Consideration}
To realise a technically feasible detector system from a conceptual design, detailed evaluations and trade-offs across multiple domains are necessary. A series of key considerations that guide the development of the high-granularity crystal ECAL are discussed below.

\subsection{Crystal Choices}\label{subsec:CrystalChoices}
The selection of the crystal material requires a comprehensive evaluation of multiple parameters, including density, scintillation characteristics, mechanical processing capability, and cost-effectiveness. A sufficiently high density is essential to achieve a short radiation length and a small Moli\`ere radius, thereby enabling a compact detector design. In addition to providing adequate light yield to meet the MIP response specifications, the crystal should have a relatively short scintillation decay time to mitigate potential signal pile-up effects. Furthermore, the crystal material needs to be suitable for manufacturing in the desired geometric dimensions and large-scale production. Practical considerations such as ease of processing, transportation, and storage also play a significant role. Given that crystal costs represent a substantial fraction of the overall ECAL expenses, minimising the unit price is also important.

Based on the aforementioned selection criteria, the BGO crystal emerges as the leading candidate. Its successful use in the ECAL of the L3 experiment~\cite{the_l3_collaboration_l3_1988} establishes a strong foundation for its potential application in the ECAL of future Higgs factories. BGO offers an advantageous combination of properties, including high density (\qty{7.13}{\gram\per\cm^3}), short radiation length (\qty{1.12}{\cm}), and relatively small Moli\`ere radius (\qty{2.26}{\cm}), which ensures effective EM shower containment. The intrinsic light yield of BGO crystals typically ranges from 8000 to \qty{10000}{\text{photons}\per\MeV}. With an optimised optical wrapping design, this light yield is sufficient to satisfy the MIP response requirements. For the mechanical properties, BGO is robust and can be easily cut and polished into long bar geometries, as evidenced by the successful production of \qty{60}{\cm} samples for the DAMPE experiment~\cite{wei_performance_2016}. The principal limitation of BGO is its relatively long scintillation decay time (\qty{300}{\ns}), which limits the achievable timing resolution and requires a correspondingly long integration time in the readout electronics. However, since the primary physics goals are centred more on energy measurement, this is an acceptable compromise.

The Lead Tungstate (PWO) crystal was successfully used in the CMS ECAL at the LHC. It features fast scintillation decay time (\qty{10}{\ns} and \qty{30}{\ns}) and high density (\qty{8.30}{\gram\per\cm^3}), which enable excellent timing performance and a compact detector design. For the present application in the high-granularity crystal ECAL, however, PWO faces certain limitations. Its intrinsically low light yield (typically 100 to \qty{200}{\text{photons}\per\MeV}) makes it difficult to achieve the targeted MIP response without high photon collection and detection efficiency. Additionally, PWO is mechanically brittle, making it extremely challenging to fabricate into the long crystal bars required by the crystal ECAL design, while also being highly susceptible to damage during transportation. These factors make PWO less suitable as the crystal material.

Other crystal options, such as Caesium Iodide (CsI) or Lutetium-yttrium Oxyorthosilicate (LYSO), were also evaluated. CsI is unsuitable for compact detector designs due to its low density, while LYSO, despite its high light yield and fast scintillation decay time, is unfeasible for large-scale applications due to the high cost.

Moreover, Bismuth Silicate (BSO) has recently gained attention as a highly promising alternative material for the crystal ECAL~\cite{deng_scintillation_2026}. With a density of \qty{6.80}{\gram\per\cm^3} and a radiation length comparable to that of BGO, BSO also ensures a compact EM shower profile. It exhibits a scintillation decay time of approximately \qty{100}{\ns}, offering improved timing performance. Although its light output is lower, roughly one-fifth to one-fourth that of BGO, preliminary studies indicate it may still be adequate to approach the targeted MIP response. The most compelling advantage is the potential for substantially reduced material cost, as silicon is far more abundant and less expensive than germanium. Recent research has made promising progress in producing large BSO samples, demonstrating its potential for the long-crystal-bar ECAL application.

In conclusion, BGO is currently the preferred baseline crystal for the ECAL. BSO presents as a potential alternative that could offer competitive performance and significantly lower cost. Ongoing research and development efforts continue to focus on fully characterising the properties of crystal materials.

\subsection{Photosensors}\label{subsec:Photosensors}
The photosensors for the crystal ECAL should achieve low noise, high sensitivity, and an exceptionally wide dynamic range spanning over five orders of magnitude to meet the detector's performance requirement, constituting a primary technical challenge.

Historically, Avalanche Photodiodes (APDs) have been successfully employed in crystal calorimeters. Their primary advantages include good linearity across a wide dynamic range, ease of integration, and insensitivity to magnetic fields. However, APDs also exhibit notable limitations for the next-generation application: they require high bias voltages (typically several hundred volts), which impose stringent demands on power supply stability to maintain gain uniformity; their moderate gain (50 to 100) limits the sensitivity to weak signals; and they suffer from excess noise along with significant temperature dependence.

SiPMs present several key features, such as operation at relatively low bias voltages (30 to \qty{50}{V}), high internal gain (\num{e5} to \num{e6}), and single-photon detection capability. These features enhance the signal-to-noise ratio and enable channel-by-channel calibration capability. Given these advantages, SiPMs have been selected as the photosensor for the crystal ECAL. To ensure compatibility with the BGO crystals, the selected SiPM device should provide a moderate Photon Detection Efficiency (PDE) that aligns with the BGO emission spectrum, while maintaining an acceptably low Dark Count Rate (DCR) to minimise noise-induced false triggers.

The principal challenge for the SiPMs lies in the dynamic range. Signals from high-energy showers can approach up to \num{4.5e5} photons per SiPM, driving the device into deep non-linearity and saturation. To mitigate this, using high-pixel-density SiPM devices with pixel pitches of \qty{6}{\um} or \qty{10}{\um} is under consideration. These products, typically with an active area of \qtyproduct[product-units=power]{3x3}{\mm}, contain hundreds of thousands of pixels, which significantly raises the saturation threshold. Moreover, the relatively long scintillation decay time of the BGO crystal helps with pixel recovery: the fired pixels can recharge and regain sensitivity, thereby extending the dynamic range. Combined with a dedicated correction algorithm, such high-pixel-density SiPMs are expected to meet the performance requirements of the crystal ECAL.

\subsection{Electronics}\label{subsec:Electronics}
The readout electronics system must accurately process signals from approximately 0.57\,M SiPM channels while meeting the requirements for dynamic range, timing resolution, power consumption, and data throughput.

Front-end electronics, implemented using the Application Specific Integrated Circuits (ASICs), face a core challenge: matching the wide dynamic range of SiPM signals. Considering a high-pixel-density SiPM with a typical gain of \num{8e4}, the signal range can span from \qty{15}{\text{p.e.}} to \qty{450000}{\text{p.e.}}. This translates to a required charge measurement range of approximately \qty{0.2}{\pico\coulomb} to \qty{5.8}{\nano\coulomb} that the ASIC must handle. To meet this demand, the ASIC design should adopt a multi-gain architecture to achieve the required energy coverage while maintaining good resolution and linearity for low-energy signals. Timing performance is another important issue in the ASIC design. Achieving the target timing resolution requires dedicated timing circuits with high bandwidth and low jitter. The preliminary design scheme of the dedicated ASIC employs a 12-bit Analogue-to-Digital Converter (ADC) and a Time-to-Digital Converter (TDC) with a resolution of \qty{100}{\ps}.

Power consumption of the front-end electronics constitutes a major engineering consideration, as the per-channel power dissipation directly governs the scale and complexity of the required cooling system. This power budget must be carefully evaluated in conjunction with specific operational parameters of the calorimeter, such as the readout scheme, data throughput, and channel occupancy, to balance the target performance with overall cooling feasibility.

The back-end electronics and Data Acquisition (DAQ) system need to handle the aggregated data flow. In the current readout architecture, data from multiple ASICs will first be aggregated by dedicated concentrator boards located at the rear of the ECAL modules, before being transmitted to the off-detector DAQ system. The back-end system should be designed with sufficient bandwidth and low-latency to manage the high-throughput data streams from all modules, while maintaining synchronisation across the entire detector and supporting potential trigger schemes.

\subsection{Mechanical Support}\label{subsec:MechanicalDesign}
The mechanical design should provide stable support for around 171 tons of BGO crystals while minimising the introduction of insensitive material. The design scheme, with front-end electronics and cooling plates integrated to the four sides of each module, pushes the material budget to the module boundaries, further compressing the available space for the support structure. Thus, the supporting framework for the crystal array itself must be extremely lightweight, compact, and rigid. 

Currently, Carbon Fibre Reinforced Polymer (CFRP) is the primary candidate for the support structure of the crystal ECAL due to its high strength and low density. A key engineering focus lies in investigating and controlling the deformation behaviour of CFRP structures under substantial mechanical loads. Detailed Finite Element Analysis (FEA) and prototyping are therefore essential to validate the geometry design and ensure structural stability.

Beyond global support, the module assembly also account for the precise alignment of crystals, which is important for the correct reconstruction of shower positions. Finally, the design of the barrel and endcap structures must facilitate easy access for module installation, replacement, and maintenance.

\subsection{Radiation Damage}\label{subsec:RadiationDamage}
Over its operational lifetime, the detector will be exposed to both ionising radiation (primarily $\gamma$) and non-ionising radiation (charged hadrons and neutrons), originating mainly from the beam-induced background and its secondary particles.

Scintillating crystals are susceptible to the Total Ionising Dose (TID) effect. When high-energy photons or charged particles deposit energy in the crystal lattice, ionisation processes generate lattice defects that give rise to colour centres. These centres enhance photon absorption, thereby reducing optical transmission and degrading light output. For BGO crystals, studies show that the light output drops rapidly at doses below \qty{e3}{\text{rad}} and remains relatively stable up to around \qty{e7}{\text{rad}}~\cite{yang_gamma-ray_2016}. The radiation damage in the BGO material requires continuous monitoring and calibration. Pre-qualification of crystal batches through irradiation tests is also essential for quality control.

SiPMs are expected to be damaged from the Non-Ionising Energy Loss (NIEL) effect. Charged hadrons and neutrons induce displacement damage in the semiconductor lattice of the SiPMs, significantly increasing the dark current and noise level, and altering the gain and photon detection efficiency of the devices. To mitigate the effects of accumulated radiation dose over long-term operation, the use of radiation-hard SiPM technologies is recommended, for example, those employing thin active layers or optimised doping profiles. Additional operation strategies may include periodic thermal annealing and compensation of the bias voltage.

Under the CEPC operating conditions, the radiation dose has been estimated based on background studies. The integrated dose is considered to be acceptable for most crystals and SiPMs within the ECAL, except for those located near the beam pipe, where the radiation damage is elevated.

\section{Performance Studies}\label{sec:Performance}
The performance of the high-granularity crystal ECAL design was investigated based on a single calorimeter module to verify its capability for achieving the $\leq3\%/\sqrt{E(\mathrm{GeV})}$ energy resolution goal. A detailed simulation framework, integrating energy deposition modelling with instrumentation effects, was employed. Sophisticated digitisation processes have been implemented to replicate the detector's realistic response.

\subsection{Simulation and Digitisation Framework}\label{subsec:SimulationDigitisation}
The full simulation of the ECAL module is based on the Geant4 toolkit, with the module geometry as described in Section~\ref{subsec:ModuleDesign}. After simulating the energy deposition within each crystal cell from incident particles, a multi-stage digitisation model is implemented to bridge the gap between this ideal response and the actual measured signals, considering the following effects in the signal generation chain:

\begin{itemize}
\item \textbf{Photon Statistics:} The energy deposition in the crystal is first converted into the number of detected photons, based on the predefined MIP response (e.g., \qty{300}{\text{p.e./MIP}}) and the known most probable MIP energy deposition in the crystal. The statistical fluctuation in photon detection is modelled by a Poisson sampling.
\item \textbf{SiPM Non-linear Response and Noise:} The response of the SiPM is simulated using a parameterised model as described in Reference~\cite{kotera_describing_2015}, accounting for their intrinsic non-linearity due to SiPM saturation and pixel recovery, as well as optical crosstalk between pixels. Subsequently, SiPM dark noise is added, modelled as a Poisson-distributed number of primary dark counts, each potentially triggering a cascade of secondary pulses via crosstalk (Borel distribution~\cite{vinogradov_analytical_2012}).
\item \textbf{Electronics Response:} The SiPM charge signal is converted to the ADC value
through the electronics response chain. This chain models the signal amplification with the gain-dependent coefficients, incorporates electronic noise via pedestal fluctuations, and accounts for potential ADC saturation by implementing an automatic gain switching mechanism. Additional ADC Integral Non-Linearity (INL) is also considered.
\item \textbf{Energy Reconstruction:} The final ADC output is converted back into the calibrated energy. This inverse process corrects for the SiPM non-linearity, subtracts the estimated dark noise contribution, and applies a calibration factor based on the MIP response. A per-channel energy threshold is then applied, and signals below this threshold are discarded.
\end{itemize}

This comprehensive digitisation model transforms the simulated ideal energy into a digitised output that closely mimics the signal that would be collected by the real detector's readout system, enabling realistic performance evaluation. The reference parameters for the crystal ECAL are listed in Table~\ref{tab:ECALDigiParameters}.

\begin{table}[htbp]
\caption{Digitisation parameters for the crystal ECAL module. These values are defined based on both the crystal ECAL design specifications and the experimental characterisations of the detector units.}  
\label{tab:ECALDigiParameters} 
\begin{tabular}{cccc}
\toprule
\textbf{Digitisation Process} & \textbf{Parameter} & \textbf{Value} & \textbf{Unit}\\ 
\midrule
\multirow{2}{*}{\textbf{Crystal}} & MIP energy deposit & 13.35  & \unit{\MeV}\\
& MIP response & 300 & \unit{\text{p.e.}\per MIP}\\
\midrule
\multirow{6}{*}{\textbf{SiPM}} & Photon detection efficiency (\qty{420}{\nm}) & 30 & \% \\
& Sensitive area & $3\times3$ & \unit{\mm^{2}}\\
& Pixel pitch & 6 & \unit{\um}\\
& Number of pixels & 244719 & --\\
& Dark count rate & 2484 & \unit{\kHz}\\
& Crosstalk & 12 & \%\\
\midrule
\multirow{10}{*}{\textbf{Electronics}} & Vertical resolution & 12 & bit\\
& Single-photon amplitude (Highest gain) & 2.5 & ADC\\
& Single-photon amplitude uncertainty & 8 & \%\\
& Pedestal position & 50 & ADC\\
& Pedestal width & 3 & ADC\\
& Number of gain stages & 3 & --\\
& Integral non-linearity & 0.5 & \%\\
\midrule
\textbf{Energy Reconstruction} & Crystal energy threshold & 0.1 & MIP\\
\botrule
\end{tabular}
\end{table}

\subsection{Energy Linearity and Resolution}\label{subsec:EMPerformance}
The module-level performance of the crystal ECAL was simulated using electrons with energies from \qty{0.1}{\GeV} to \qty{100}{\GeV}, incident perpendicularly onto the centre of the module. For each event, the total energy was reconstructed by summing the digitised energy deposition from all detector cells within the module. 

\begin{figure}[htbp]
\centering
\includegraphics[width=0.7\textwidth]{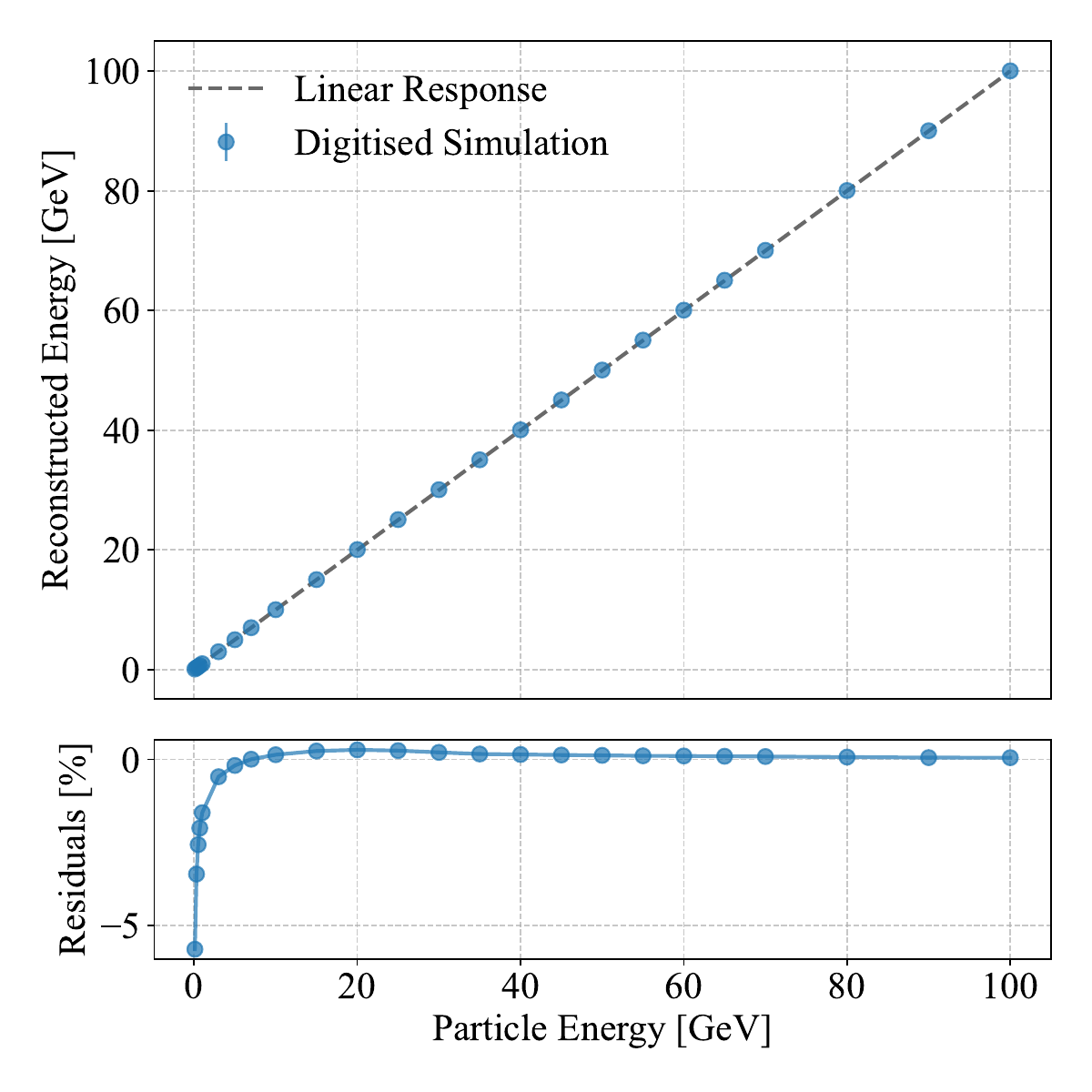}
\caption{Energy linearity of the reconstructed energy for electrons incident on a crystal ECAL module. The deviation from perfect linearity is within $\pm0.5\%$ across the \qtyrange{3}{100}{\GeV} range.}\label{fig:CrystalModuleEnergyLinearity}
\end{figure}

Figure~\ref{fig:CrystalModuleEnergyLinearity} presents the energy linearity of the crystal ECAL module. Over the primary energy range from approximately \qtyrange{3}{100}{\GeV}, the reconstructed energy exhibits excellent linearity with respect to the incident electron energy, with deviations remaining within $\pm0.5\%$. At lower energies, significant non-linearity is observed, primarily due to the energy threshold applied to each channel. This low-energy non-linearity can be effectively corrected through dedicated calibration procedures.

\begin{figure}[htbp]
\centering
\includegraphics[width=0.7\textwidth]{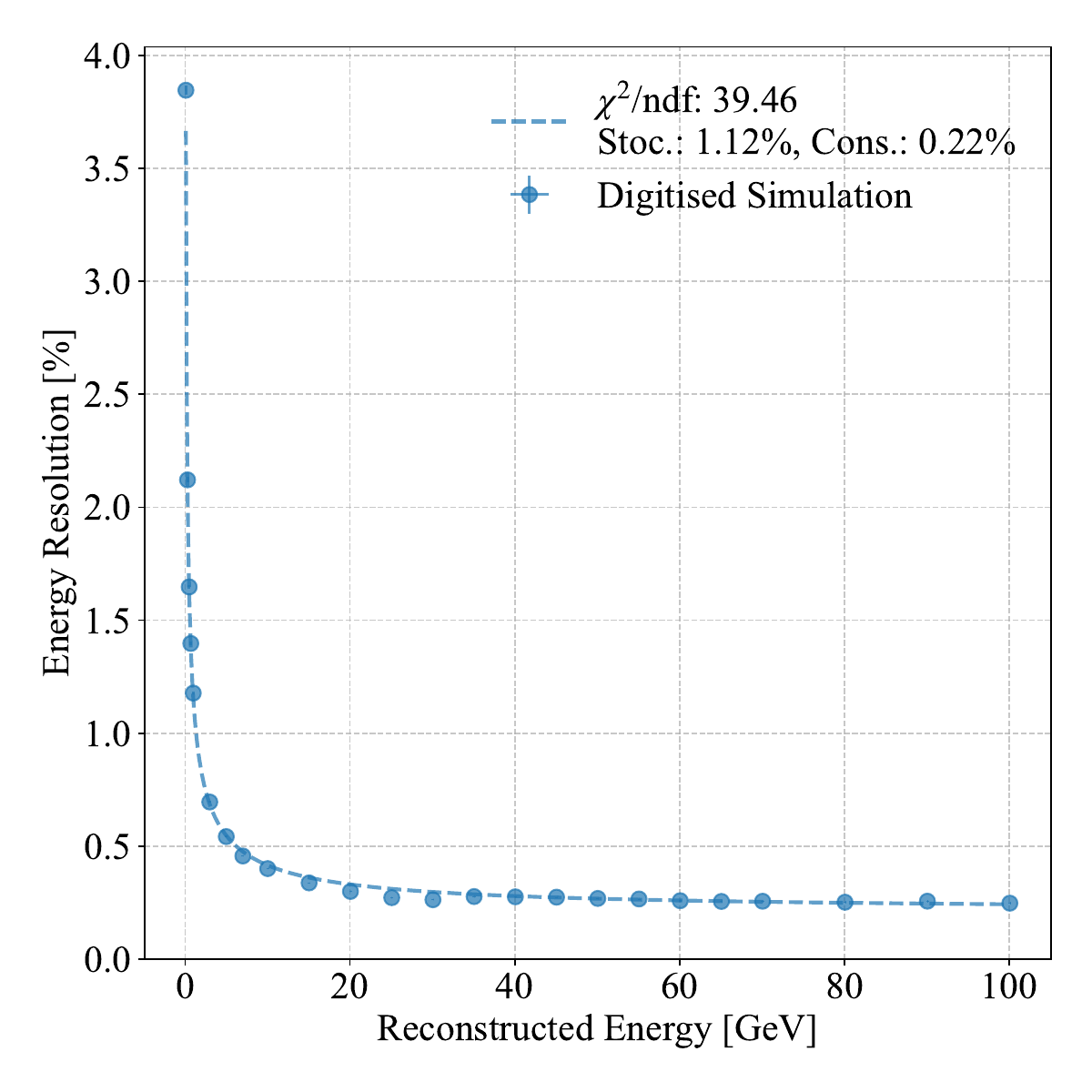}
\caption{Energy resolution of a single crystal ECAL module as a function of incident electron energy. The fitted stochastic and constant terms are 1.12\% and 0.22\%, respectively.}\label{fig:CrystalModuleEnergyResolution}
\end{figure}

The electromagnetic energy resolution is shown in Figure~\ref{fig:CrystalModuleEnergyResolution}. The achieved resolution is parameterised as $\sigma_{E}/E=1.12\%/\sqrt{E(\mathrm{GeV})}\oplus0.22\%$, which not only meets but significantly exceeds the design goal. This result demonstrates the superior intrinsic capability for precise energy measurement.

These results demonstrate that the single-module performance of the high-granularity crystal ECAL design far exceeds the requirement of $\leq3\%/\sqrt{E(\mathrm{GeV})}\oplus1\%$. The achieved performance provides strong evidence that the proposed concept successfully combines the excellent intrinsic energy resolution of a homogeneous calorimeter with a finely segmented geometry suitable for PFA. While this study focuses on an ideal, isolated module, the significant performance margin offers a comfortable buffer for the full detector system. Future work will focus on quantifying the system-level effects through more comprehensive simulations and prototype studies.

\section{Summary and Prospects}\label{sec:Summary}
The conceptual design of a novel high-granularity crystal ECAL for future lepton colliders has been presented. Driven by the stringent physics requirement for a jet energy resolution capable of delivering a 4\% level boson mass resolution, the proposed ECAL synergises the superior intrinsic energy resolution of a homogeneous calorimeter with the fine segmentation necessary for particle-flow reconstruction. Its core design employs orthogonally arranged long scintillating crystal bars read out by SiPMs at both ends.

The development of the ECAL module and full-detector geometry, alongside a dedicated particle-flow software, has been outlined. Through comprehensive studies, key design specifications have been established, and critical technical considerations across detector subsystems have been discussed. 

Module-level performance evaluations, with a detailed digitisation model, demonstrate that the design can achieve an EM energy resolution of $1.12\%/\sqrt{E(\mathrm{GeV})}\oplus0.22\%$ with excellent linearity ($\pm0.5\%$ from \qtyrange{3}{100}{\GeV}), confirming the feasibility of reaching its performance goal. The development of the module, full detector geometry and the particle-flow software has been outlined. Key design specifications have been established through detailed studies. Critical design considerations have been discussed.

Looking ahead, several key issues require focused research and development to advance the design towards technical realisation. A primary challenge lies in the dynamic range of front-end electronics and SiPMs. This necessitates the selection of high-pixel-density SiPMs with optimised saturation behaviour, and the design of readout ASICs with multi-gain architectures to accommodate the required wide charge range. The PFA reconstruction software requires further enhancement to address the unique pattern recognition challenges presented by the module geometry, including ambiguity removal and shower separation. The mechanical design requires continued investigation into lightweight, high-strength structures to support the substantial crystal mass while minimising insensitive material.

Besides, the development of robust and precise calibration schemes represents a crucial task, particularly to account for factors such as temperature variations and radiation-induced damage in the detector. A comprehensive calibration strategy will be essential, including pre-installation characterisation of crystal-SiPM units, in-situ calibration of electronic pedestals and noise levels, and physics-based methods utilising well-understood processes such as Bhabha scattering, $Z\to e^{+}e^{-}$ and $\pi^{0}\to\gamma\gamma$ decays. By combining these approaches, the calibration procedures enable per-channel energy scaling, ensuring that the performance target can be met and maintained over the full operational lifetime of the detector.

Further important areas for future work include investigating radiation damage effects on detector components and developing radiation-hard SiPMs and crystals. Additionally, research into alternative crystal materials, such as BSO, should be continued to explore options for performance optimisation and potential cost reduction.

The high-granularity crystal ECAL represents a promising technological pathway toward the precision calorimetry required at next-generation lepton colliders. Ongoing and future research efforts will be essential to transform this conceptual design into a state-of-the-art detector system.

\backmatter

\bmhead{Acknowledgements}
The authors would like to express their sincere gratitude to all colleagues of the CEPC calorimetry working group for their collaborative efforts and helpful discussions. This research was supported by the National Key Research and Development Program of China (Grant Number 2023YFA1606300).

\section*{Declarations}
\begin{itemize}
\item \textbf{Conflict of interest:} The authors declare that there is no conflict of interest.
\end{itemize}


\bibliography{sn-bibliography}

\end{document}